\documentclass[copyright,creativecommons]{eptcs}
\providecommand{\event}{FMAS 2026}
\def\titlerunning{From Attack Tree to Model Checker: Binding CAPEC Patterns}
\def\authorrunning{J. Hussain, L. A. Dennis, C. Dixon \& M. Farrell}

\usepackage{breakurl}
\usepackage{booktabs}
\usepackage{listings}
\usepackage{xcolor}
\usepackage{amsmath,amssymb}
\usepackage{graphicx}
\usepackage{url}
\usepackage{enumitem}
\usepackage{verbatim}
\usepackage{booktabs,tabularx}

\definecolor{codebg}{HTML}{F6F8FA}
\definecolor{codeline}{HTML}{D0D7DE}
\lstdefinestyle{ltl}{backgroundcolor=\color{codebg}, basicstyle=\ttfamily\scriptsize, frame=single, rulecolor=\color{codeline}, framesep=4pt, breaklines=true, columns=fullflexible, keepspaces=true, showstringspaces=false, aboveskip=5pt, belowskip=5pt}
\AtBeginDocument{\setlength{\abovedisplayskip}{5pt}\setlength{\belowdisplayskip}{5pt}\setlength{\abovedisplayshortskip}{3pt}\setlength{\belowdisplayshortskip}{3pt}}
\newcommand{\pre}{\mathit{pre}}
\newcommand{\eff}{\mathit{eff}}
\newcommand{\Th}{\theta}
\newcommand{\G}{\Box}
\newcommand{\Fin}{\Diamond}
\newcommand{\U}{\,\mathcal{U}\,}
\newcommand{\pred}[1]{\ensuremath{\mathrm{#1}}}
\newcommand{\ent}[1]{\ensuremath{\mathit{#1}}}
\newcommand{\TS}{\mathrm{TS}}

\usepackage[colorinlistoftodos]{todonotes}

\newcommand{\marienote}[1]{\todo[inline, color=magenta!65, author=Marie]{#1}}

\title{From Attack Tree to Model Checker:\\ Binding CAPEC Patterns to Verify Robotic System Security\\[3pt]
}
\author{Juel Hussain \qquad Louise A. Dennis \qquad Clare Dixon \qquad Marie Farrell
\institute{Department of Computer Science, University of Manchester, UK}
}

\begin{document}
\maketitle

\begin{abstract}
Security-minded verification of autonomous and robotic systems presupposes an attacker, yet the
path from a threat catalogue to a checkable attacker model is still traversed by hand. This research
preview presents the theory and construction of a pipeline that mechanises it. Each leaf of an attack
tree is mapped to a CAPEC attack pattern and written in first-order logic as a rule: the attacker’s pre-
requisites imply the attack’s effect. Casting the pattern as such a rule is what makes it operational - its variables can be bound to a specific system’s parts, and its prerequisites can then be tested against that system under a chosen configuration of security controls. Binding is our means of verifying whether a system is vulnerable to a given attack: from a bound pattern we derive an attacker model, turn its effect into a temporal-logic property, and check that property against the system with and without a countermeasure, reporting the control that defeats the attack. We describe the signature, schemas, profiles for approach, and explore how CAPEC can support reasoning about robotics security.
\end{abstract}

\section{Introduction}
\label{sec:intro}

Formal verification of autonomous robotic systems has primarily focused on safety concerns~\cite{Luckcuck2019Survey,Dennis2022Verifiable,Cardoso2022Compositional}; however, there is a growing need to adapt to security properties in the context of an attacker~\cite{Farrell2019SEFM,Maple2020Space}. 
In our previous work~\cite{Hussain2026ABZ} on autonomous satellite docking, the composed attacker-system model was manually constructed and verified in SPIN~\cite{Holzmann2003SPIN} against security properties expressed in LTL. Although the approach was rigorous, the manual approach was time-consuming and does not scale. This has motivated the use of Attack Trees~\cite{Schneier1999,MauwOostdijk2005} (as they enumerate, in a structured way, how an adversary might reach a goal) and CAPEC~\cite{CAPEC} (catalogues several hundred named attack patterns). However, neither is a formal object and today an analyst bridges the gap between them and a model checker by hand. This research preview presents a way that mechanises this bridge. 
We introduce a pipeline that uses two logics which do different jobs. A CAPEC pattern is written once, for no particular system, so it needs variables: a first-order action schema over a sorted signature says what an attacker must already have, and what changes when the action succeeds. A system deployment is a finite set of ground atoms: a static description of the system. Binding instantiates the schema against the deployment. It turns a catalogue entry into a decidable claim about a system under a named configuration of security controls, and when the claim fails it reports the specific condition responsible, which identifies the control that defeats the attack. Instantiation also fixes a finite propositional vocabulary, and the grounded actions generate an attacker transition system over it. The transition preserves the grounded schema by using its prerequisites as the transition guard and its effect as the transition update; the temporal property then asks whether that same effect is reachable. Only at that point does temporal logic become the right tool: LTL states properties of the runs of the grounded attacker model composed with the system model, $\TS_{sys}\parallel_{ch}\TS_{att}$, in the input-mediated form of~\cite{Hussain2026ABZ}. This is the mechanised form of the \emph{Input-Mediated Interaction} (IMI) composition. This is illustrated in Fig.~\ref{fig:composition}. A CAPEC pattern does not become part of the system model itself: after binding, it induces a separate attacker transition system which interacts with the system model only through the identified system interfaces.


\paragraph{Contributions.} A formal grounding construction that transforms catalogue-level attack knowledge into system-specific attacker models suitable for compositional model checking. Supported by (i) CAPEC operational semantics: a many-sorted action-schema representation (Section \ref{sec:inputs}); (ii) System grounding: a binding relation between CAPEC schema parameters and robotic-system entities, parameterised by security control configurations (Section \ref{sec:binding});  (iii) Verification: construction of attacker transition systems and temporal properties from grounded attack patterns for composition with system models (Section \ref{sec:ltl}).

\section{Background \& Related Work}
\label{sec:background}

An attack tree~\cite{Schneier1999} defines a root goal through OR and AND nodes to \emph{leaves}, which are the basic attacker actions to achieve the sub-goal (nodes). Mauw~\cite{MauwOostdijk2005} gave the standard semantics (a set of multisets of leaves and the attack suites), Jhawar et al.~\cite{Jhawar2015SAND} added sequential conjunction (SAND) which provides order in sub-goals to achieve overall goal. We use OR, AND and SAND in our trees with the suite semantics of per-leaf results to whole-tree properties~\cite{MauwOostdijk2005}. MITRE's CAPEC~\cite{CAPEC} catalogues natural-language attack patterns, each with an abstraction level, \emph{prerequisites}, \emph{execution flow} (optional) and \emph{consequences} of security attacks. Of the 558 non-deprecated patterns in CAPEC (v3.9), 349 (62.5\%) have non-empty \emph{Prerequisites} and \emph{Consequences}. Both are needed for a schema; former is needed as a precondition to trigger on, and the latter is needed as an effect to assert. This allows expressiblity over the vocabulary (predicate signature of capability, state, progress, and violation predicates) of our inputs. A further 239 (42.8\%) carry an execution flow, the source of multi-step schemas. These counts bound what any vocabulary can formalise: the 349 are the patterns a schema can be written for at all, while whether a given one is expressible over our signature is decided pattern by pattern (Table~\ref{tab:capec-examples}). We use first-order logic in many-sorted form, so every name in the model has a declared type (e.g. a camera is a sensor and not a channel), over a fixed signature - list of allowed predicates and types are settled in advance. Table~\ref{tab:capec-examples} gives an expressible and an inexpressible pattern. 
A system is a finite set of ground atoms (factual statements with no variables left e.g. camera input is authenticated) taken as all the true atoms (everything that holds of the system) so under the closed-world assumption (anything absent from the list is taken to be false, rather than unknown), entailment of ground literals (deciding whether a given specific fact, or its negation, follows from system description) is membership testing. This is the discipline behind MulVAL's Datalog (a closed-world logic restricted in the same way)~\cite{Ou2005MulVAL}. Properties are LTL~\cite{Pnueli1977} ($\G$, $\Fin$, $\U$; SPIN's \texttt{[]}, \texttt{<>}, \texttt{U} in listings) checked with SPIN~\cite{Holzmann2003SPIN}, which returns a counterexample trail on violation. Existential questions (``can attacker reach $s$?") are emitted as $\G\neg s$ and a violation is read as reachability, the trail as witness. Our attacker models follow the IMI construction of~\cite{Hussain2026ABZ}: an attacker transition system composed with the system at a designated input channel, $\TS_{sys}\parallel_{ch}\TS_{att}$.

\begin{table}[t]
\centering\small
\begin{tabular}{@{}p{0.20\linewidth} p{0.24\linewidth} p{0.09\linewidth} p{0.39\linewidth}@{}}
\toprule
CAPEC & Description & Expressible & Schema/Reason \\
\midrule
148-Content Spoofing &
  Modifies content so that it no longer says what its provider intended &
  Yes \newline 
  &
  $\textsf{CanModify}(a,r)\wedge\textsf{Carries}(ch,r)\wedge\neg\textsf{Authenticated}(ch)\rightarrow\textsf{IntegrityViolated}(r)$ \\
25-Forced Deadlock &
  Triggers a deadlock in the target software to deny service, e.g.\ a ROS~2
  node blocked on a synchronous service call that never returns &
  No &
  Not a premise: both the deadlock condition and its effect are properties of the
  system's runs, not facts about the deployment. Checkable once the attacker is
  composed with a model of the system. \\
\bottomrule
\end{tabular}
\caption{Expressible and inexpressible CAPEC patterns. Schema for 148 is in AT2FOL~\cite{AT2FOL}.}
\label{tab:capec-examples}
\end{table}

\medskip
\noindent\textbf{Related Work:} Audinot et al.~\cite{Audinot2017Correct} label nodes with pre-/post-conditions over a transition system and define when a refinement is admissible, consistent and complete. This is the closest work to our relevance check. Their conditions are hand-authored per node and the focus is the tree's structural correctness, while our work focuses on the applicability of a bound pattern to a deployment under named controls, produced from a public catalogue via an explicit binding, which is then handed to a model checker. Other work synthesises trees from a system model~\cite{Pinchinat2014ATSyRA}. 
MulVAL~\cite{Ou2005MulVAL,Ou2006Scalable} derives multi-stage attack reachability from scanner-reported host and vulnerability facts using exploit rules written by experts in Datalog. Our applicability check uses the same closed-world fragment, but its rules come from the catalogue: each CAPEC pattern is written once as a schema and reused across systems, and the output is not a reachability graph but, for each control configuration, a verdict naming the control that blocks the attack, together with an attacker process and LTL properties for SPIN~\cite{Ammann2002Scalable}. Kumar et al.~\cite{Kumar2015PTA} model-check attack trees for quantitative metrics. 
Farrell et al.~\cite{Farrell2019SEFM} and Maple et al.~\cite{Maple2020Space} select properties with STRIDE~\cite{Shostack2014} and verify them in Promela/SPIN and Dafny; we mechanise the step from threat to property that they perform by hand, with CAPEC in place of STRIDE. Unlike STRIDE's six threat categories, a CAPEC pattern supplies the structure needed to derive precondition/effect schemas and decide applicability under a control configuration mechanically.

\section{Security Verification Mechanisation: Architecture and Overview}
\label{sec:pipeline}

\begin{figure}[t]\centering
\includegraphics[width=0.9\textwidth]{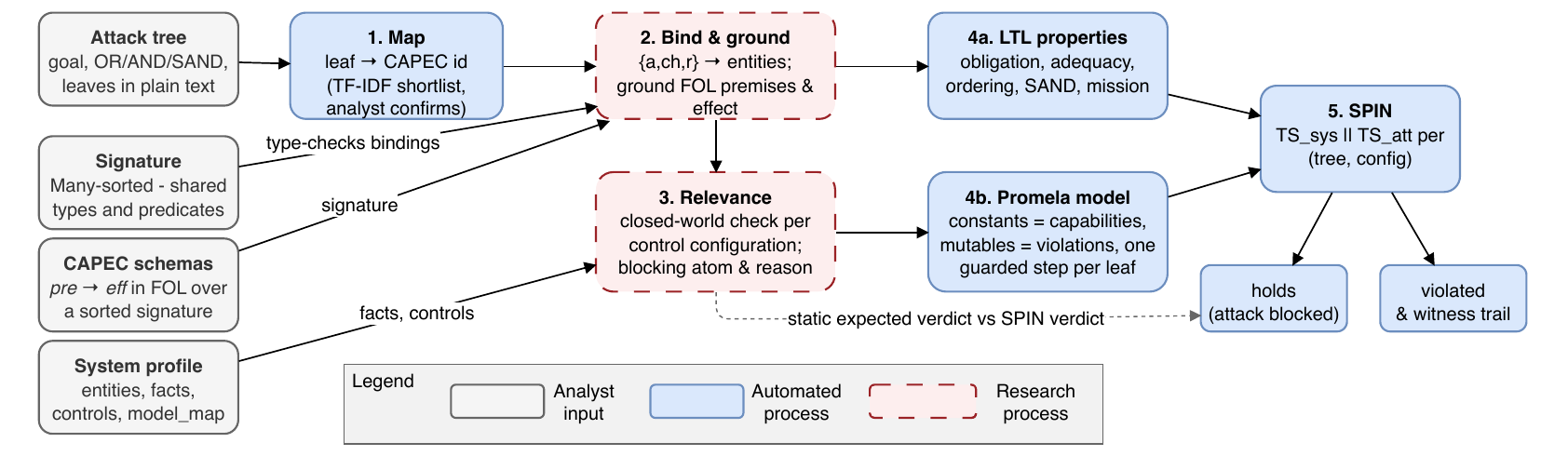}
\caption{The Security Verification Mechanisation pipeline. Four authored inputs (signature, schemas, profile, and the attack tree) feed a mapping step, the binding-and-relevance step at the centre, and the emission of LTL properties and a Promela attacker that SPIN checks per configuration. The static relevance verdict is compared with SPIN's.}
\label{fig:pipeline}
\end{figure}

Fig.~\ref{fig:pipeline} shows the pipeline, implemented in the prototype AT2FOL~\cite{AT2FOL}\footnote{\url{https://github.com/juelhussain/AT2FOL}, release \texttt{fmas2026} (work in progress).}. Analysts author the four inputs, confirm CAPEC mappings and supply bindings. All subsequent stages are mechanically generated; the construction constraints imposed on these artefacts are therefore critical to the validity of the generated verification models. 

\subsection{Building the inputs: signature, schemas, profiles, and trees}
\label{sec:inputs}

The pipeline takes four inputs, deliberately kept independent: a signature that fixes the vocabulary, schemas that formalise CAPEC patterns without reference to any system, profiles that describe a deployment without reference to any attack, and trees that scope which attacks are considered.

The signature fixes the sorts and predicates in which every artefact; schemas, profiles, properties, must be written so that a CAPEC pattern authored independently of any system can still be grounded against a particular deployment. Without that shared vocabulary there is nothing to bind against: ``the camera feed is unauthenticated" in a schema and in a profile would be unrelated strings rather than the same atom, and the relevance check would have no fact to test. Every artefact is written over the single many-sorted signature of Fig.~\ref{fig:signature-profile}, whose predicates fall into four classes: capability, state, progress and violation. Capability and state record what a deployment can do and how it is wired up; progress and violation record what an attacker gains mid-attack and what a completed attack causes. A profile may assert only the first two, and a schema step may produce only the last two.

\begin{figure}[t]
\begin{lstlisting}
# signature.yaml (excerpt) - the vocabulary shared by every schema, profile and property
sorts:      [ Agent, Component, Channel, Message, Resource ]
capability: [ CanObserve(Agent,Channel), CanTransmit(Agent,Channel), CanModify(Agent,Resource),
              Authenticated(Channel), Encrypted(Message), RateLimited(Channel) ] # what a deployment can do
state:      [ Carries(Channel,Resource), OnChannel(Message,Channel),
              InUse(Channel), Consumes(Component,Message) ]   # how the system is wired up
progress:   [ KnowsProtocol(Agent,Channel), Positioned(Agent,Channel),
              HasSession(Agent,Channel), KnowsFormat(Agent,Resource) ]  # gained part-way through
violation:  [ IntegrityViolated(Resource), ConfidentialityFlag(Resource),
              AvailabilityViolated(Resource), AccessControlViolated(Resource)] # caused by an attack
              
# capec-148.yaml - Content Spoofing, written for no particular system 
params:  { a: Agent, ch: Channel, r: Resource }
pre:      [ "CanModify(a,r)", "Carries(ch,r)", "not Authenticated(ch)" ]
eff_add:  [ "IntegrityViolated(r)" ]
eff_del:  [ ]
    
# satellite-docking.yaml (excerpt) - one deployment over that vocabulary
entities:   { Channel: [gnss_in, cam_in, cmd_in, rf_uplink],
              Resource: [nav_solution, pose_estimate, dock_authority, command_link] }
facts:      [ "CanModify(attacker, pose_estimate)", "Carries(cam_in, pose_estimate)",
              "CanTransmit(attacker, rf_uplink)", "Carries(rf_uplink, command_link)", ... ]
controls:   { cam_channel_auth:  {adds:    ["Authenticated(cam_in)"]},
              uplink_isolation:  {removes: ["CanTransmit(attacker, rf_uplink)"]},
              uplink_rate_limit: {adds:    ["RateLimited(rf_uplink)"]}, ... }
configurations: [ {id: C1, controls: []}, {id: C2, controls: [cam_channel_auth]}, ..., {id: C6, ...} ]
model_map:  { channels:  { cam_in: {promela_chan: cam_chan, guard_var: cam_ok}, ... },
              atom_vars: { "IntegrityViolated(pose_estimate)": pose_integrity_violated, ... } }
\end{lstlisting}
\caption{The three artefacts that must share a vocabulary. The signature fixes the sorts and predicates; the CAPEC-148 schema is written for no particular system, and the satellite-docking profile for no particular attack. A profile asserts capability and state atoms only, and a schema step produces progress and violation atoms only: the profile says what the system is, the schema says what an attack changes. Only the binding $\Th = \{a \mapsto \ent{attacker},\ ch \mapsto \ent{cam\_in},\ r \mapsto \ent{pose\_estimate}\}$ brings them together.}
\label{fig:signature-profile}
\end{figure}

A schema is the operational reading of one CAPEC pattern: what the catalogue describes in prose, the schema states as conditions and effects. Formally it comprises sorted parameters (placeholders such as ``some channel", instantiated later against a deployment), a precondition (a conjunction of literals over the parameters) and an effect (atoms added or deleted); multi-step patterns carry a list of such steps, each step's effect supplying facts to the next step. Example schemas have been authored such as CAPEC 148 (Content Spoofing), 125 (Flooding), 94 (Adversary in the Middle), and 157 (Sniffing Attacks). A system profile describes one deployment: entities per sort, which are the actual channels, resources and components of the system under analysis together with a closed-world set of base facts (capability and state atoms only), named \emph{controls} that add or remove ground atoms to model a defence being switched on or off, and \emph{configurations} bundling controls, each configuration standing for one hardening level of the same system. The satellite-docking profile from~\cite{Hussain2026ABZ} is captured in Fig.~\ref{fig:signature-profile}. The \texttt{model\_map} ties each channel to its Promela channel and guard variable and each violation atom to a readable name.

\begin{figure}[t]\centering
\includegraphics[width=0.75\textwidth]{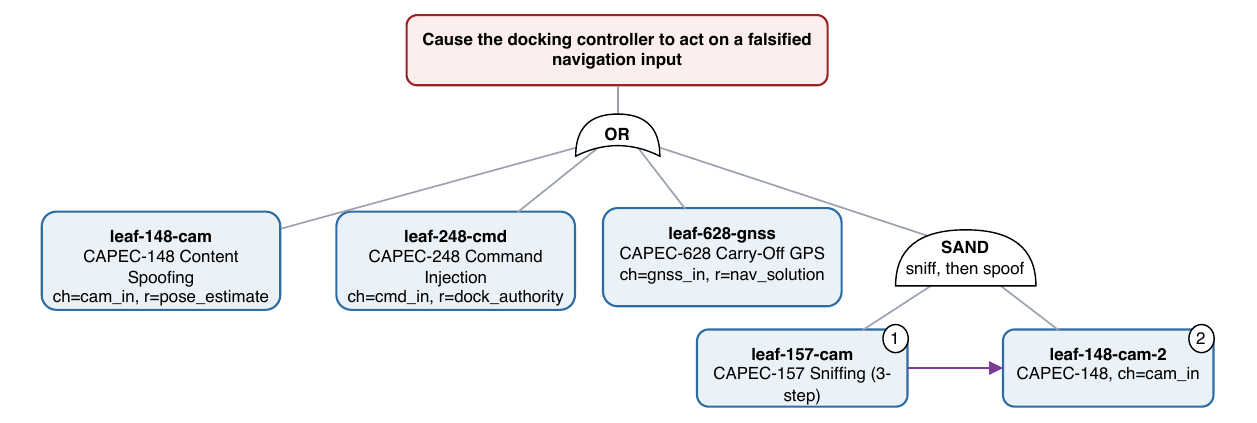}
\caption{The satellite-docking attack tree with its SAND refinement. Every leaf carries a CAPEC identifier and a binding of the schema's parameters $(a,ch,r)$ to profile entities. Numbers show sequence.}
\label{fig:tree}
\end{figure}

We took a collection of attack-tree leaf descriptions and paired them with the CAPEC pattern that we consider to be the best match (goldset). 
The trees contain alternative, combined, and ordered attack paths using OR, AND, and SAND refinements. Fig.~\ref{fig:tree}, for example, presents a satellite-docking scenario in which several alternative attacks may achieve the overall goal, including an ordered path where the attacker first observes a camera channel and then spoofs it.

\medskip
\noindent\textbf{From leaves to CAPEC to FOL.} Given a leaf's text the tool ranks the CAPEC catalogue (558 patterns; names weighted threefold) by TF--IDF cosine similarity and returns a shortlist with the literal matched terms, so every score is inspectable ~\cite{Kanakogi2021Information}. 
Mapping is therefore \emph{analyst-assisted}: the tool offers a shortlist, and the analyst confirms. 
A CAPEC entry says in prose what an attacker needs and what happens if they succeed; the schema says it as a rule. For parameters $\bar{x}=(a{:}\mathsf{Agent},\,ch{:}\mathsf{Channel},\,r{:}\mathsf{Resource})$, representing the attacking agent, communication channel, and affected resource, respectively, a single-step schema is $\forall\bar{x}.\ \pre(\bar{x})\rightarrow\eff(\bar{x})$, with $\pre$ a conjunction of literals over capability, state and progress predicates and $\eff$ a conjunction of violation (or progress) atoms. Three of the schemas, CAPEC-125 (Flooding), CAPEC-148 (Content Spoofing) and CAPEC-158/651 (Sniffing Network Traffic and Eavesdropping), read:
{\small\begin{align*}
\text{CAPEC-125:}\quad&
\pred{CanTransmit}(a,ch)\wedge\pred{Carries}(ch,r)\ \rightarrow\ \pred{AvailabilityViolated}(r)\\
\text{CAPEC-148:}\quad&
\pred{CanModify}(a,r)\wedge\pred{Carries}(ch,r)\wedge\neg\pred{Authenticated}(ch)\ \rightarrow\ \pred{IntegrityViolated}(r)\\
\text{CAPEC-158/651:}\quad&
\pred{CanObserve}(a,ch)\wedge\pred{Carries}(ch,r)\ \rightarrow\ \pred{ConfidentialityFlag}(r)
\end{align*}}
\noindent
The \emph{negated} literal $\neg\pred{Authenticated}(ch)$ is how a pattern says ``only on an unauthenticated channel'', exactly the literal a channel-authentication control falsifies. The map from CAPEC patterns to schemas can be many-to-one. CAPEC-158 and CAPEC-651 share the same schema, despite being different attacks, because observation is handled as a single capability ($\pred{CanObserve}(a,ch)$). But CAPEC separates the attacks by medium, i.e. eavesdropping does not take place on a network channel while sniffing does. Distinguishing the medium would require a finer signature, for example a predicate marking network channels, which we leave to future work. Multi-step patterns are chains whose later premises read a \emph{progress} atom an earlier step produced. CAPEC-157 (Sniffing Attacks), the Standard-level parent of CAPEC-158, is the only one of the three whose entry has an execution flow, which gives such a chain:
{\small\begin{align*}
&\pred{CanObserve}(a,ch)\rightarrow\pred{KnowsProtocol}(a,ch);\qquad
\pred{KnowsProtocol}(a,ch)\rightarrow\pred{Positioned}(a,ch);\\
&\pred{Positioned}(a,ch)\wedge\pred{Carries}(ch,r)\rightarrow\pred{ConfidentialityFlag}(r)
\end{align*}}%

\noindent
Each step's effect enables the next: observing the channel yields protocol knowledge. This enables the attacker to position itself on the channel, from where it obtains the information carried by that channel. The chain form is what later yields ordering and partial-progress properties: progress atoms are the only place a temporal dependency between attacker actions is \emph{stated} rather than assumed.

\subsection{Binding and relevance: is this system vulnerable to this attack?}
\label{sec:binding}

\begin{figure}[t]\centering
\includegraphics[width=0.9\textwidth]{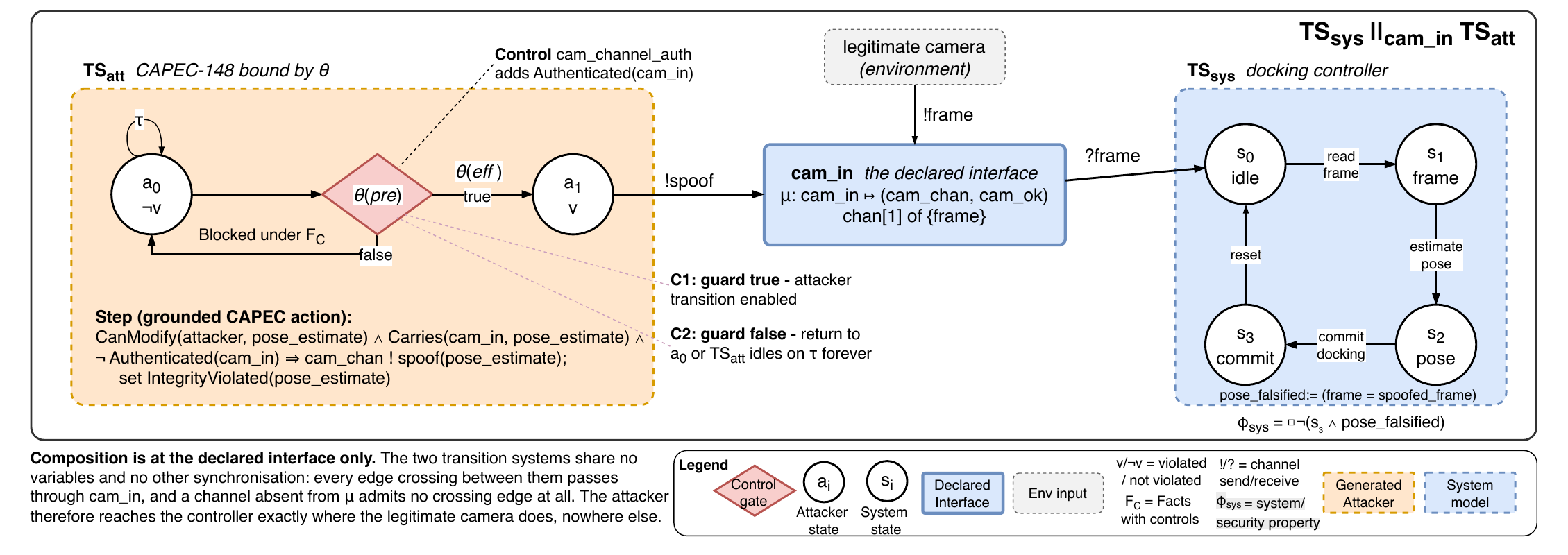}
\caption{The composition for the camera case. $\mathrm{TS}_{att}$ is the transition system induced by the grounded CAPEC-148 rule; $\mathrm{TS}_{sys}$ is the docking controller. They share no variables: every crossing edge passes through \texttt{cam\_in}, the interface \texttt{model\_map} declares, and \texttt{cam\_channel\_auth} acts on that interface, falsifying the guard under $C2$. $\G\neg\mathrm{IntegrityViolated}(\ent{pose\_estimate})$ mentions only atoms
$\mathrm{TS}_{att}$ writes, so relevance decides it alone; $\G\neg(s_3 \wedge \ent{pose\_falsified})$ mentions $\mathrm{TS}_{sys}$'s own state and is decidable only here. 
}
\label{fig:composition}
\end{figure}

A \emph{binding} $\Th$, supplied by the leaf, maps a schema's parameters to profile entities of matching sort; a binding is rejected if the sorts disagree or if it names a channel absent from \texttt{model\_map}, since an undeclared interface offers no composition point. For
\texttt{leaf-148-cam},
$\Th=\{a\mapsto\ent{attacker},\ ch\mapsto\ent{cam\_in},\ r\mapsto\ent{pose\_estimate}\}$, so
$\Th(\pre)=\pred{CanModify}(\ent{attacker},\ent{pose\_estimate})\wedge
\pred{Carries}(\ent{cam\_in},\ent{pose\_estimate})\ \wedge\neg\pred{Authenticated}(\ent{cam\_in})$
and $\Th(\eff)=\pred{IntegrityViolated}(\ent{pose\_estimate})$. Thus, the grounded action is enabled when the attacker can modify the pose estimate carried by the unauthenticated camera channel; executing it establishes $\pred{IntegrityViolated}(\ent{pose\_estimate})$. The same schema bound differently is a distinct grounded attack instance. Fig.~\ref{fig:composition} shows this action synchronising with the controller only through \ent{cam\_in}, where \ent{cam\_channel\_auth} disables its guard. 

For each configuration $C$, let $F_C$ be the facts holding once its controls have been applied. A bound action is \emph{applicable} in $C$ when every positive literal of $\Th(\pre)$ is in $F_C$ and every negated one is absent. Multi-step chains are checked in order, effects carried forward. When a literal fails we report it as the \emph{blocking atom}, with reason \texttt{missing} / \texttt{present\_but\_negated}, so a blocked verdict points to the control that defeats the attack. The camera leaf is applicable under $C1$ and blocked under $C2$, where \ent{cam\_channel\_auth} establishes $\pred{Authenticated}(\ent{cam\_in})$ and so falsifies the premise's negated literal, reported \texttt{present\_but\_negated}. 
Verdicts are also relative to profile completeness: an unrecorded capability makes an attack look blocked, so we report ``blocked under $F_C$", never ``secure".

\subsection{From bound rules to LTL and SPIN}
\label{sec:ltl}

A grounded rule yields the attacker and its properties in one move. Atoms become Promela variables via \texttt{model\_map}; capability and state atoms are constants of the configuration, progress and violation atoms mutable and initially false. Each step becomes one guarded \texttt{atomic} branch of an attacker process, with a $\tau$ branch letting the attacker decline to act. Each emitted property carries a \emph{polarity} recording how to read SPIN's verdict: a violation of a universal claim means the property fails, whereas a violation of an existential one means the target is reachable and the trail is the witness. With $v_1, \dots, v_k$ the leaf's grounded violation atoms, $p_i=\Th(\pre_i)$ the grounded precondition of step~$i$ and $d_i$ its completion, we emit the templates of Table~\ref{tab:ltl}: per leaf, an \emph{obligation}, \emph{adequacy} guards that detect vacuity, \emph{partial progress} and step \emph{ordering}; per tree, an ordering claim for each SAND node and a \emph{reachability} claim for each attack suite. The listing shows the instances emitted for the tree of Fig.~\ref{fig:tree}.

\begin{lstlisting}[style=ltl]
ltl leaf_148_cam_obligation { [] !pose_integrity_violated }
ltl leaf_148_cam_adequacy   { [] !(can_modify_attacker_pose_estimate && carries_cam_in_pose_estimate
                                   && !authenticated_cam_in) }
ltl leaf_157_step2_partial  { [] !(knows_protocol_attacker_cam_in && [] !pose_confidentiality_flag) }
ltl leaf_157_cam_step1_partial { [] !(can_observe_attacker_cam_in
      && [] !pose_confidentiality_flag) }
ltl sand_sniff_then_spoof   { ([] !pose_integrity_violated) || (!pose_integrity_violated U pose_confidentiality_flag) }
\end{lstlisting}

\begin{table}[t]
\centering\footnotesize
\begin{tabularx}{\textwidth}{@{}p{2.7cm} p{2.7cm} X l X@{}}
\toprule
Property & Formula & Intended claim & Polarity & SPIN counterexample means \\
\midrule
Obligation & $\Box\neg(v_1\vee\dots\vee v_k)$ & the attack's effect never occurs
  & universal & the attack succeeds; the trail is the attack \\
Adequacy & $\Box\neg p_n$ & the final step's precondition is reachable
  & existential & it is reachable, so the obligation is not vacuous \\
Step adequacy ($i<n$) & $\Box\neg p_i$ & step $i$ is reachable
  & existential & step $i$ is reachable \\
Ordering ($i\ge 2$) & $\Box\neg d_i \vee{}$\newline $(\neg d_i\,\mathcal{U}\,d_{i-1})$
  & step $i$ never completes before step $i-1$ & universal & a step can be skipped \\
Partial progress ($i<n$) & $\Box\neg(p_i\wedge\Box\neg d_n)$
  & some run reaches step $i$ and never completes the chain & existential
  & such a run exists \\
\midrule
SAND ordering & $\Box\neg\mathit{done}(l_2)\vee(\neg\mathit{done}(l_2)\, \mathcal{U}\,\mathit{done}(l_1))$
  & leaf $l_2$ never completes before $l_1$ & universal & the tree's order is not enforced \\
\bottomrule
\end{tabularx}
\caption{LTL templates emitted per leaf (top) and per tree (bottom).
$\mathit{done}(l)$ is the conjunction of leaf $l$'s grounded violation atoms.}
\label{tab:ltl}
\end{table}

SPIN runs once per (tree, configuration, property), checking the emitted obligation, adequacy, partial-progress and ordering properties over the composition $\TS_{sys}\parallel_{ch}\TS_{att}$. We plan to run every configuration in turn showing which control discharges which leaf.

\section{Evaluation Plan \& Conclusion}
\label{sec:concl}
There are four dimensions to the evaluation. We evaluate on CAPEC retrieval effectiveness and extend against ATT\&CK-BERT and other embeddings for comparability with~\cite{Kanakogi2021Information,Birchler2024CRiSIS}. We assess schema coverage, static grounding to look at expected applicability vs actual applicability and finally evaluate the formal generation and look at generated SPIN result vs expected/manual model. 


We have presented, as a research preview, a pipeline that takes an attack tree, attaches a CAPEC pattern to each leaf, writes it as a first-order rule, binds the rule to a system profile, decides under each control configuration whether the bound attack applies and which control defeats it, and emits the LTL and Promela attacker that SPIN checks. Binding is the centre: where a catalogue pattern becomes a verifiable claim about a deployment, and the mechanised form of the input-mediated attacker composition of our earlier hand-built verification. The pipeline addresses (i) Applicability: Is the CAPEC attack applicable to the system/configuration? (ii) Reachability: Can the instantiated attacker achieve its CAPEC-defined effect? (iii) Vulnerability: Does the attacker cause a violation of a security property of the actual system? In our previous work~\cite{Hussain2026ABZ} we established $\TS_{sys}\parallel_{ch}\TS_{att}$ for a manually authored attacker. We are now defining a pipeline that generates the attacker model $G(CAPEC,\theta,F_C)=TS_{att}$ and so \(TS_{sys} \parallel_{ch} G(CAPEC,\theta,F_C)\) will be verified.

\bibliographystyle{eptcs}
\bibliography{refs}
\end{document}